\documentclass[aps,prl,preprint,superscriptaddress]{revtex4-2}

\usepackage{graphicx}  
\usepackage{dcolumn}   
\usepackage{bm}        
\usepackage{amsmath}   
\usepackage{amssymb}   
\usepackage{booktabs}  
\usepackage{hyperref}  
\usepackage{array}     

\newcommand{\WHon}{WHon}

\begin{document}

\title{WH Statistics: Generalized Pauli Principle for Partially Distinguishable Particles}

\author{Wang Hao}
\email{25b905039@stu.hit.edu.cn}
\author{Meng Yancen}
\author{Zhang Kuang}
\email{zhangkuang@hit.edu.cn}
\thanks{Corresponding author.}
\affiliation{School of Electronics and Information Engineering, Harbin Institute of Technology, Harbin 150001, China}

\author{Zhou Rui'en}
\affiliation{School of Mathematics, Harbin Institute of Technology, Harbin 150001, China}

\date{\today}

\begin{abstract}
Traditional statistical mechanics is constrained by the binary paradigms of identical/distinguishable and bosonic/fermionic particle statistics, resulting in a fundamental logical gap in describing systems with partial distinguishability. 
We propose the WH Statistics, a unified theoretical framework governed by continuous distinguishability $\lambda$, exclusion weight $\kappa$, and intrinsic exclusivity $\gamma$. 
By deriving the microstates and entropy, we demonstrate that the proposed framework naturally recovers the Bose-Einstein, Fermi-Dirac, and Maxwell-Boltzmann limits, while incorporating anyons and the classical hard-core (Langmuir) limit. 
We introduce a class of generalized quasiparticles, termed \WHon s, which exhibit exotic physical phenomena—including degeneracy pressure peaks, Schottky-like specific heat anomalies, and other characteristic effects—driven by the interplay of fractional distinguishability and exclusivity.
This framework bridges the century-old gap between quantum and classical exclusion, offering a powerful tool for investigating strongly correlated systems.
\end{abstract}


\maketitle

The standard binary paradigm of statistical mechanics leaves a critical gap in the distinguishable but strongly exclusive quadrant, often erroneously reducing it to the Maxwell-Boltzmann limit\cite{LandauStatPhys}. This oversight renders the current framework incapable of describing classical hard-core (Langmuir) systems. In this work, we establish Wang-Hao (WH) statistics. Our contributions are threefold. First, we introduce the \textit{continuous distinguishability} and \textit{exclusivity weight} hypotheses to resolve the compatibility between continuous parameters and discrete configuration counting. Second, we prove that WH statistics unifies classic quantum distributions and anyons while uniquely incorporating the Langmuir system ($\lambda=1, \kappa=1$) into a single paradigm (Table 1\ref{tab:WH_systems}). Third, we define the \WHon\ quasiparticle family and predict falsifiable signatures, including non-monotonic degeneracy pressure, Schottky-like specific heat anomalies, and interference effects tunable by fractional distinguishability.

\begin{table*}[t]
    \caption{Particle unifiable under the WH Statistics}  
    \label{tab:WH_systems}  
    \begin{ruledtabular}  
    \begin{tabular}{l c c c l}
        \textbf{System} & $\boldsymbol{\lambda}$ & $\boldsymbol{\kappa}$ & $\boldsymbol{\gamma}$ & \textbf{Ref.} \\  
        \colrule
        Ideal Boson                     & $0$           & $0$           & $0$           & \cite{Bose1924,Einstein1925,Anderson1995} \\
        Fermion                         & $0$           & $1$           & $1$           & \cite{Fermi1926,Dirac1926,Pauli1925} \\
        Tonks-Girardeau (TG) gas        & $0$           & $1$           & $0$           & \cite{Girardeau1960,Kinoshita2004} \\
        Anyon                           & $0$           & $\alpha \in (0,1)$ & $0$           & \cite{Wilczek1982,Haldane1991,Wu1994,Bartolomei2020,Zhang2025} \\
        Classical Ideal Gas             & $1$           & $0$           & $0$           & \cite{LandauStatPhys} \\
        Langmuir System                 & $1$           & $1$           & $0$           & \cite{Langmuir1917} \\
        \textbf{Generalized WHon Family} & $(0, 1]$     & $(0, 1]$     & $0, 1$        & --- \\
    \end{tabular}
    \end{ruledtabular}
\end{table*}  

To construct a statistical mechanical description, we introduce two core axioms based on experimental phenomenology. First, we treat particle \textit{indistinguishability} as a continuous degree of freedom $\lambda \in [0,1]$, defined microscopically by the wavefunction overlap $\lambda \equiv |\langle \psi_i | \psi_j \rangle|^2$. This parameter smoothly bridges the classical distinguishable limit ($\lambda \to 0$) and the strict identical quantum limit ($\lambda \to 1$). Experimentally, $\lambda$ maps directly to the visibility $V$ in generalized Hong-Ou-Mandel (HOM) interference experiments via the scaling relation $V \propto \lambda^2$ \cite{Hong1987,Mandel1991,Bocquillon2013,Shchesnovich2015,Ra2013,Yousef2025}.

Second, we quantify the competition between quantum exclusion and thermal fluctuations via the \textit{effective exclusivity weight} $\kappa \in [0,1]$. We propose a Padé-approximant evolution equation linking the microscopic interaction strength $U$ and temperature $T$:
\begin{equation}
\kappa(\mathcal{X}) = \frac{\mathcal{X}}{\mathcal{X} + \xi}, \quad \text{with} \quad \mathcal{X} \equiv \frac{U}{\sqrt{k_B T}},
\label{eq:kappa_evolution}
\end{equation}
where $\mathcal{X}$ serves as the dimensionless competition parameter and $\xi$ is a geometry-dependent constant (e.g., $\xi \approx \hbar/\sqrt{m}$ for 1D $\delta$-potentials). This ansatz captures the crossover from thermal dominance ($\mathcal{X} \ll 1, \kappa \to 0$) to interaction-dominated hard-core blocking ($\mathcal{X} \gg 1, \kappa \to 1$).

To complete the framework, we introduce the discrete topological parameter $\gamma \in \{0, 1\}$ to characterize intrinsic exchange symmetry, where $\gamma=0$ denotes symmetric wavefunctions (integer spin) and $\gamma=1$ denotes antisymmetric ones (half-integer spin). Combining this with the aforementioned axioms, the unified microscopic configuration number $\Omega_{\text{WH}}$ for $N$ particles occupying $G$ states is constructed as:
\begin{equation}
\Omega_{\text{WH}} = \frac{1}{(N!)^\lambda} \frac{\big[G + (1-\kappa)(1-\gamma)(N-1)\big]!}{\big[G - \kappa N - (1-\kappa)(1-\gamma)\big]!}.
\label{eq:microstates}
\end{equation}

In the thermodynamic limit ($N, G \gg 1$), applying Stirling's approximation yields the generalized WH entropy:
\begin{equation}
S_{\text{WH}} \approx k_B \Big[ \mathcal{A} \ln \mathcal{A} - \mathcal{B} \ln \mathcal{B} - \lambda (N \ln N - N) \Big],
\label{eq:entropy}
\end{equation}
where $\mathcal{A} \equiv G + \tilde{\kappa} N$ and $\mathcal{B} \equiv G - \kappa N$, with $\tilde{\kappa} \equiv (1-\kappa)(1-\gamma)$. Crucially, the term $-\lambda k_B (N \ln N - N)$ acts as a tunable Gibbs correction factor; it naturally recovers the factor $1/N!$ in the quantum identical limit ($\lambda \to 1$) and vanishes in the classical limit ($\lambda \to 0$), thereby resolving the mixing entropy paradox within a continuous manifold.

Maximizing $S_{\text{WH}}$ under particle and energy conservation yields the implicit transcendental equation for the average occupation number $n_{\text{WH}} = N/G$:
\begin{equation}
\beta(\epsilon - \mu) = \ln \left( \frac{(1 + \tilde{\kappa} n_{\text{WH}})^{\tilde{\kappa}} (1 - \kappa n_{\text{WH}})^{-\kappa}}{n_{\text{WH}}^\lambda} \right).
\label{eq:transcendental}
\end{equation}
This solution can be parameterized into the unified WH distribution function,
\begin{equation}
n_{\text{WH}}(\epsilon) = \frac{1}{e^{\beta(\epsilon - \mu)} + \Theta(\kappa, \gamma, n_{\text{WH}})},
\label{eq:wh_dist}
\end{equation}
where the generalized statistical function $\Theta$ encapsulates the correction due to partial distinguishability and exclusion:
\begin{equation}
\Theta(\kappa, \gamma, n) \equiv \frac{1}{n} \left[ 1 - n^{1-\lambda} (1 + \tilde{\kappa} n)^{\tilde{\kappa}} (1 - \kappa n)^{\kappa} \right].
\label{eq:theta}
\end{equation}
Eqs.~(\ref{eq:transcendental})--(\ref{eq:theta}) constitute the analytical core of WH statistics, enabling a continuous transition from quantum identical statistics to classical localized statistics within a single framework.
The WH distribution accurately recovers standard statistical limits at specific asymptotic boundaries. For identical particles ($\lambda=1$), the interplay of exclusion and symmetry naturally bifurcates: fermions ($\kappa=1, \gamma=1$) force $\Theta \to 1$, yielding the Fermi-Dirac distribution $n_{\text{FD}} = (e^{\beta(\epsilon-\mu)} + 1)^{-1}$, while bosons ($\kappa=0, \gamma=0$) lead to $\Theta \to -1$, reproducing the Bose-Einstein distribution $n_{\text{BE}} = (e^{\beta(\epsilon-\mu)} - 1)^{-1}$. In the regime of fractional exclusion ($\lambda=1, 0 < \kappa < 1$), the dilute limit yields linear scaling $\Theta \approx 2\kappa - 1$, recovering Haldane-Wu statistics \cite{Wilczek1982,Haldane1991,Wu1994} which continuously bridges Bose condensation and Fermi blocking.

Crucially, the framework resolves the microscopic origin of the classical hard-core limit. By tuning to the classical distinguishable regime ($\lambda \to 0$) with steric repulsion ($\kappa=1$), Eq.~(\ref{eq:wh_dist}) yields the Langmuir distribution:
\begin{equation}
n_{\text{Lang}} = \frac{1}{e^{\beta(\epsilon - \mu)} + 1} = \frac{K(T) P}{1 + K(T) P},
\label{eq:langmuir}
\end{equation}
where we utilize the fugacity relation $e^{\beta \mu} \propto P$ and define the adsorption constant $K(T)$. This derivation proves that the unity term in the denominator arises strictly from classical steric exclusion rather than quantum antisymmetry, clarifying a foundational concept in surface science \cite{Langmuir1917}.

\begin{figure}[t]
    \centering
    \includegraphics[width=0.9\columnwidth]{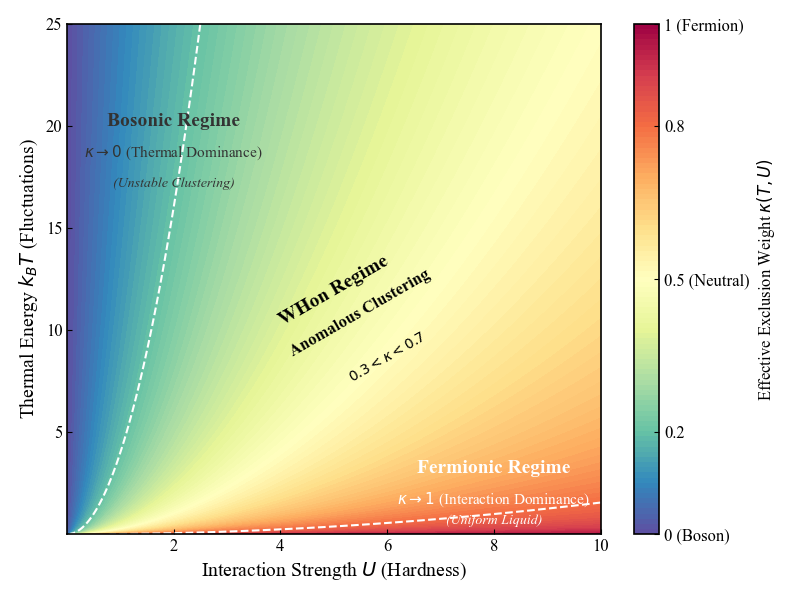} 
    \caption{Phase diagram of the WHon regime. The color gradient represents the effective exclusion weight $\kappa(T,U)$. The WHon state emerges in the crossover region (yellow) between the unstable bosonic clustering (blue, $\kappa \to 0$) and the rigid fermionic/hard-core limit (red, $\kappa \to 1$), stabilized by the interplay of interaction strength $U$ and thermal energy $k_B T$.}
    \label{fig:phase_diagram}
\end{figure}

Beyond these limiting cases, the framework predicts the existence of the WHon quasiparticle, specifically in the regime of strong exclusion ($\kappa \to 1$) while retaining partial distinguishability ($0 < \lambda < 1$). The stability region of this quasiparticle is mapped in Fig.~\ref{fig:phase_diagram}, defined by the competition between interaction hardness $U$ and thermal fluctuations.

We identify two physical realizations. First, recent work on spin-depairing mechanisms in non-Hermitian systems reveals exotic superfluid phase transitions driven by Exceptional Points (EPs) \cite{Takemori2025}. In the WH framework, the spin-depairing rate $\gamma$ acts as a decoherence source ($\lambda \propto 1 - \gamma/\gamma_c$), driving the system from a pure coherent state to a mixed state. The competition between attractive Hubbard pairing ($U<0$) and residual fermionic hard-core nature ($\kappa \to 1$) maps the superfluid EPs directly onto the critical points of the WH phase diagram. Second, within the screening cloud formed around magnetic impurities ($T < T_K$) \cite{Kondo1964,Hewson1993}, electrons acquire partial distinguishability ($\lambda < 1$) due to impurity tagging, while strong Coulomb repulsion enforces hard-core exclusion ($\kappa \approx 1$). Treating the Kondo cloud as a WHon gas naturally explains the effective mass divergence without ad hoc hybridization assumptions.

\begin{figure}[b]
    \centering
    \includegraphics[width=\columnwidth]{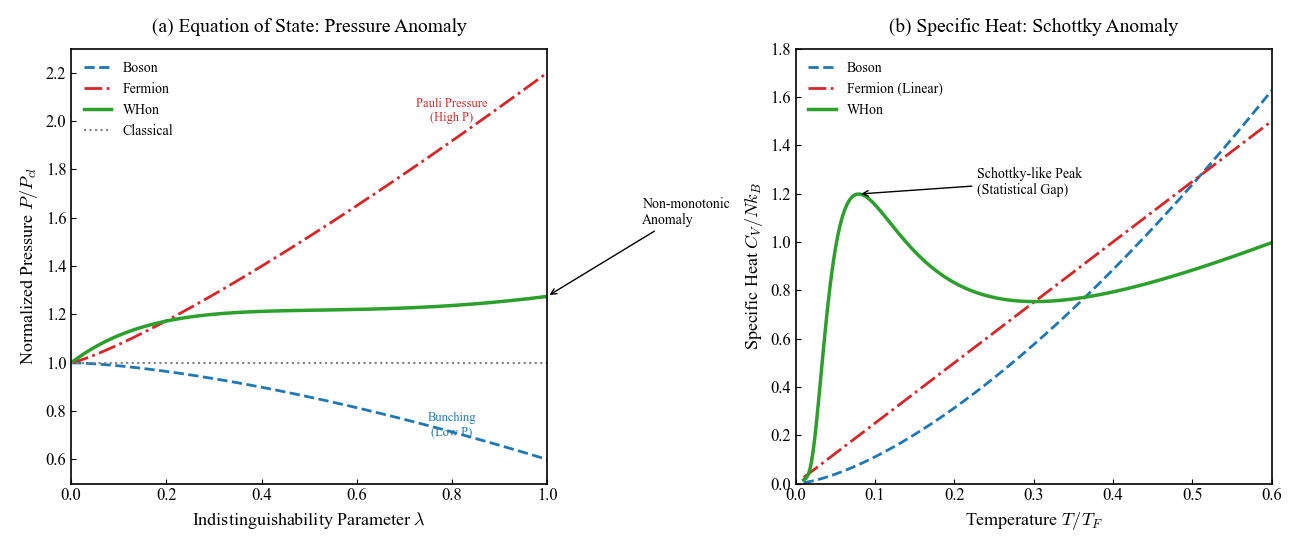} 
    \caption{Thermodynamic anomalies of WHon fluids. (a) Normalized degeneracy pressure $P/P_{cl}$ versus indistinguishability $\lambda$. Unlike Bosons (blue dashed) or Fermions (red dot-dashed), the WHon equation of state (green solid) exhibits a non-monotonic pressure peak. (b) Isochoric specific heat $C_V$. The exclusion-induced statistical gap leads to a Schottky-like anomaly at low temperatures, distinct from the linear behavior of Fermi liquids.}
    \label{fig:thermo}
\end{figure}

Theoretical analysis indicates four distinguishing features of WHon fluids. First, the thermodynamics is governed by a non-monotonic pressure anomaly arising from the competition between entropy gain (driven by $\lambda$) and phase space blocking (driven by $\kappa$). Unlike monotonic Fermi systems, the WHon pressure exhibits a local maximum at a critical distinguishability $\lambda_c \approx (1-\kappa)/\kappa$ [see Fig.~\ref{fig:thermo}(a)]. This peak arises when the sublevel population hits the hard-core boundary ($N/M > 1/\kappa$), forcing a statistical rearrangement that serves as a definitive thermodynamic criterion. 

Second, the isochoric specific heat $C_V$ possesses a statistical gap induced by $\kappa$, contrasting with gapless Fermi liquids. In the ultra-low temperature regime ($T \approx 0.05 T_F$), we predict a Schottky-like peak [Fig.~\ref{fig:thermo}(b)] where the peak temperature follows the scaling $T_{\text{peak}} \propto \kappa / \ln(1/\lambda)$, providing a spectroscopic method to extract microscopic parameters.

\begin{figure}[t]
    \centering
    \includegraphics[width=\columnwidth]{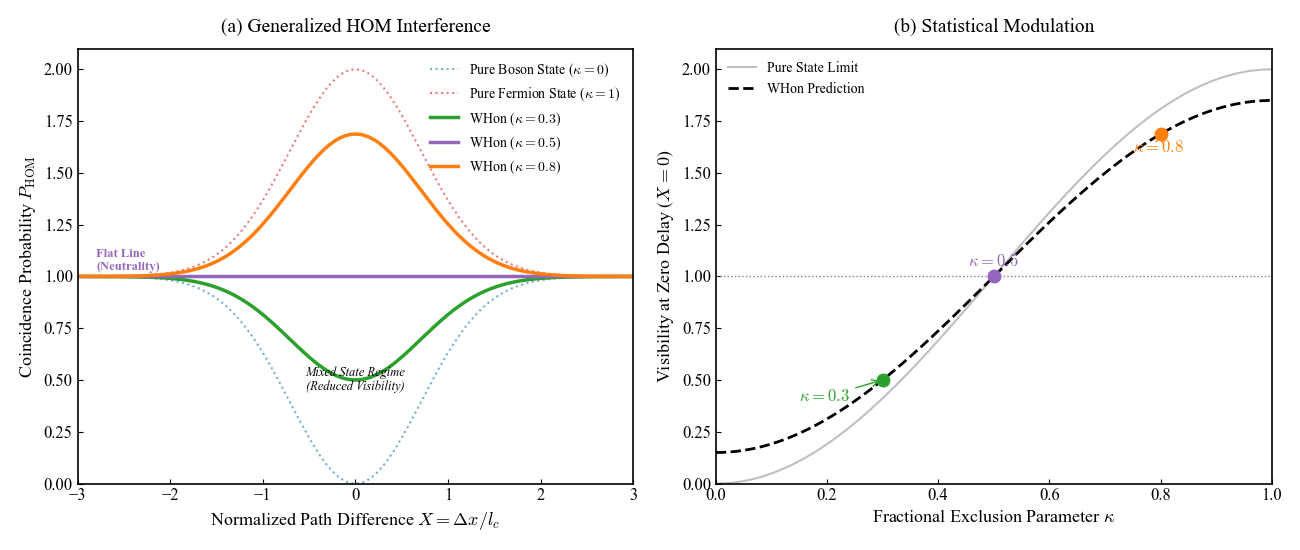} 
    \caption{Generalized HOM interference signatures. (a) Coincidence probability profiles. The WHon state (green/orange lines) exhibits intermediate bunching/anti-bunching behavior tuned by $\kappa$. (b) Statistical modulation of visibility at zero delay. The dashed line represents the WHon prediction, unifying the pure Bose limit ($\kappa=0$) and Fermi limit ($\kappa=1$).}
    \label{fig:HOM}
\end{figure}

Third, interference patterns offer a striking signature, being jointly modulated by fractional distinguishability and exclusion. We derive the generalized Hong-Ou-Mandel (HOM) probability:
\begin{equation}
P_{\text{HOM}}(\Delta x, \kappa) = 1 - \exp\left[ - \left( \frac{\Delta x}{\beta \lambda_{\text{dB}}} \right)^2 \right] \cos(\pi \kappa).
\label{eq:hom_whon}
\end{equation}
As shown in Fig.~\ref{fig:HOM}, the temporal width is limited by geometric decoherence ($\lambda_{\text{dB}}$), while the saturation amplitude is governed by $\kappa$. This ``statistically constrained decoherence'' suggests that recent anyon collision experiments \cite{Bartolomei2020} function essentially as HOM interferometers, where interference signals may vanish even under finite overlap at statistical critical points.

Finally, these macroscopic properties manifest spatially as fractional anti-clustering. WHons adhere to a ``Soft Blockade'' principle: while finite $\lambda$ permits spatial overlap, the repulsion from $\kappa \to 1$ enforces long-range anti-correlations to minimize thermodynamic potential, creating a statistically forbidden region $0 < g^{(2)}(0) < 0.5$. Near the critical pressure point $\lambda_c$, the correlation function transitions from exponential to power-law decay, stabilizing the exceptional fermionic superfluid state.

In summary, the WH Statistics constructs a logically complete paradigm that bridges the discontinuity between quantum and classical statistics, formally resolving the Gibbs paradox. By interpreting the continuous distinguishability parameter $\lambda$ as an ensemble average over discrete rational configurations, we reconcile macroscopic continuity with integer microstate counts. This formulation predicts that while thermodynamic quantities evolve smoothly in the thermodynamic limit, they will manifest observable stepwise transitions in finite-size systems. Furthermore, we establish that the exclusion weight $\kappa$ is not a phenomenological fitting parameter but a rigorous order parameter anchored in the ratio of interaction energy to thermal fluctuations, whereas the intrinsic symmetry $\gamma$ remains strictly discrete due to topological constraints of the 3D permutation group. Beyond recovering standard distributions, the prediction of \WHon\ quasiparticles provides decisive experimental signatures. Ultimately, this unified framework equips the community with a precise theoretical tool for engineering the thermodynamics of programmable quantum matter at the interface of partial distinguishability and strong correlations.

\bibliography{sample} 
\bibliographystyle{apsrev4-2} 

\end{document}